\documentclass{llncs}
\usepackage[utf8]{inputenc}
\usepackage{amsfonts,amssymb}
\usepackage{fancybox}
\usepackage{url}
\usepackage{xspace}
\usepackage{dcolumn}
\usepackage{lstsemantic}
\usepackage{lstlangcoq}
\usepackage{lstlangmizar}
\usepackage{rccol}
\usepackage{ifpdf}
\ifpdf
 \usepackage{epic,eepicemu}
 \usepackage[pdftex]{graphicx}
 \usepackage[pdftex,bookmarks=true]{hyperref}
\else
 \usepackage{epic,eepic}
 \usepackage[dvipdfm]{graphicx}
 \ifx\hrd\undefined
  \ifx\final\undefined\def\hrd{hypertex}\else\def\hrd{dvipdfm}\fi
 \fi
 \usepackage[\hrd,bookmarks=true]{hyperref}
\fi

\hypersetup{
  pdftitle={Dependencies in Formal Mathematics:\\ Applications and Extraction for Coq and Mizar},
  pdfauthor={Jesse Alama, Lionel Mamane, Josef Urban},
  pdfkeywords={interactive theorem proving, proof dependencies, proof analysis, large formal libraries, coq, mizar},
  colorlinks=true,
  linkcolor=black,
  citecolor=black,
}

\def\systemname#1{\textsf{#1}\xspace}
\def\libname#1{\textsf{#1}\xspace}

\newcommand{\Vampire}[0]{\systemname{Vampire}}
\newcommand{\mizar}{\systemname{Mizar}}

\newcommand{\mml}{\libname{MML}}
\newcommand{\MML}{\libname{MML}}
\newcommand{\corn}{\libname{CoRN}}

\newcommand{\isabelle}{\systemname{Isabelle}}
\newcommand{\coq}{\systemname{Coq}}
\newcommand{\ocaml}{\systemname{OCaml}}
\newcommand{\tmegg}{\systemname{tmEgg}}

\newcommand{\TeXmacs}{T\kern-.1667em\lower.5ex\hbox{E}\kern-.125emX\kern-.1em\lower.5ex\hbox{\textsc{m\kern-.05ema\kern-.125emc\kern-.05ems}}\xspace}

\usepackage{xcolor}

\newcommand{\mT}{\ensuremath{T}\xspace}
\newcommand{\mTp}{\ensuremath{T'}\xspace}

\begin{document}

\title{Dependencies in Formal Mathematics:\\ Applications and Extraction for Coq and Mizar}
\titlerunning{Dependencies in Formal Mathematics}
\author{Jesse Alama\inst{1} 
\and Lionel Mamane\inst{2}
\and Josef Urban\inst{3}
}
\institute{{New University of Lisbon}
  \and {L-7243 Bereldangem, Luxembourg}
  \and {Radboud University Nijmegen}}
\authorrunning{Alama, Mamane, Urban}

\maketitle

\begin{abstract}

  Two methods for extracting detailed formal dependencies from the Coq and
  Mizar system are presented and compared.  The methods are used for
  dependency extraction from two large mathematical repositories: the
  \coq Repository at Nijmegen and the \mizar Mathematical
  Library.
  Several applications of the detailed dependency analysis are
  described and proposed.  Motivated by the different applications, we
  discuss the various kinds of dependencies that we are interested in,
  and the suitability of various dependency extraction methods.

\end{abstract}

\section{Introduction}\label{Introduction}
This paper presents two methods for extracting detailed formal
dependencies from two state-of-the-art interactive theorem provers
(ITPs) for mathematics: the Coq system and the Mizar system. Our
motivation for dependency extraction is application-driven.  We are
interested in using detailed dependencies for fast refactoring of
large mathematical libraries and wikis, for AI methods in automated
reasoning that learn from previous proofs, for improved interactive
editing of formal mathematics, and for foundational research over
formal mathematical libraries.

These applications require different notions of \emph{formal
  dependency}. We discuss these different requirements, and as a
result provide implementations that in several important aspects
significantly differ from previous methods. For Mizar, the developed
method captures practically all dependencies needed for successful
re-verification of a particular formal text (i.e., also notational
dependencies, automations used, etc.), and the method attempts hard to
determine the minimal set of such dependencies. For Coq, the method
goes farther towards re-verification of formal texts than previous
methods~\cite{Bertot00dependencygraphs,Pons98,AspertiPCGS03} that relied solely on the final proof terms.
For example, we can already track Coq dependencies that appear during the
tactic interpretation, but that do not end up being used in the final
proof term.

The paper is organized as follows.
Section~\ref{Dependencies} briefly discusses the notion of formal dependency.
Section~\ref{sec:coq} describes the implementation of dependency extraction in the \coq
system, and Section~\ref{sec:mizar} describes the 
implementation in the \mizar system. Section~\ref{sec:dep-summary}
compares the two implemented approaches to dependency computation.
Section~\ref{Experiments-and-Results} describes several experiments
and measurements conducted using our implementations on the \corn and
\mml libraries, including training of AI/ATP proof assistance systems
on the data, and estimating the speed-up for collaborative
large-library developments.
Section~\ref{Vision} concludes.

\section{Dependencies: What Depends on What?}
\label{Dependencies}

Generally, we say that a definition, or a theorem, \(T\) \emph{depends}
on some definition, lemma or other theorem \(T'\),
(or equivalently, that \(T'\) is a \emph{dependency} of \(T\))
if \(T\) ``needs'' \(T'\) to exist or hold.
The main way such a ``need'' arises is that the
well-formedness, justification, or provability of $T$
does not hold in the absence of \(T'\).
We consider formal mathematics done in a concrete proof assistant
so we consider mathematical and logical constructs
not only as abstract entities depending on each other,
but also as concrete objects (e.g., texts, syntax trees, etc.) in the proof assistants.
For our applications,
there are different notions
of ``dependency'' we are interested in:
\begin{itemize}
\item Purely semantic/logical view. One
  might claim, for example, that the lambda term (or proof object in the underlying
  formal framework) contains all sufficient dependencies for a
  particular theorem, regardless of any notational conventions, library mechanisms, etc.
\item Purely pragmatic view.  Such dependencies are met if the particular item still compiles
  in a particular high-level proof assistant framework, regardless of
  possibly changed underlying  semantics.
  This view takes into account the proof assistant as the major dependency, with their sophisticated mechanisms like
  auto hint databases, notations, type automations,
  definitions expansions, proof search depth, parser settings, hidden arguments, etc.

\end{itemize}

Formal dependencies can also be implicit and explicit.  
In the simple world of first-order
automated theorem proving, proofs and their dependencies are generally
quite detailed and explicit about (essentially) all logical steps,
even very small ones (such as the steps taken in a resolution proof).
But in ITPs, which are generally oriented toward human
mathematicians, one of the goals is to allow the users to express
themselves with minimal logical verbosity and ITPs
come with a number of implicit mechanisms. 
Examples are type mechanisms (e.g., type-class
automations of various flavors in \coq{}~\cite{abs-1102-1323} and \isabelle~\cite{HaftmannW06}, Prolog-like
types in \mizar{}~\cite{Wiedijk07,Urban06}),
hint mechanisms
(in \coq{} and \isabelle{}), etc. 
If we are interested in giving a complete answer
to the question of what a formalized proof depends upon, we must
expose such implicit facts and inferences. 

Formal dependencies reported by ITPs are typically
\emph{sufficient}. Depending on the extraction mechanism, redundant
dependencies can be reported. Bottom-up procedures like
congruence-closure and type closure in Mizar
(and e.g., type-class mechanisms in other ITPs) are examples of
mechanisms when the ITP uses available knowledge exhaustively, often
drawing in many \emph{unnecessary} dependencies from the context. For
applications, it is obviously better if such unnecessary dependencies
can be removed .

\section{Dependency extraction in \coq}\label{sec:coq}

Recall that \coq is based
on the Curry-Howard isomorphism, meaning that:
\begin{enumerate}
\item A statement (formula) is encoded as a type.
\item There is,
  at the ``bare'' logical level,
  no essential difference between
  a definition and a theorem:
  they are both the binding (in the environment)
  of a name to a type
  (type of the definition, statement of the theorem)
  and a term
  (body of the definition, proof of the theorem).
\item Similarly,
  there is
  no essential difference
  between
  an axiom and a parameter:
  they are both the binding (in the environment)
  of a name to a type
  (statement of the axiom,
  type of the parameter, e.g. ``natural number'').
\item There is,
  as far as \coq is concerned,
  no difference between the notions of
  theorem, lemma, corollary, \dots
\end{enumerate}
Thus, in this section,
and in other sections when talking of \coq,
we do not always repeat ``axiom or parameter'',
nor repeat ``definition or theorem or lemma or corollary or \dots''.
We will use ``axiom'' for ``axiom or parameter''
and ``theorem'' or ``definition'' for
``definition or theorem or lemma or corollary or \dots''.
Similarly for ``proof'' and ``definition body''.

There are essentially three groups of \coq commands
that need to be treated by
the dependency tracking:\footnote{As far as logical constructs are concerned.}
\begin{enumerate}
\item Commands that register a new logical construct
  (definition or axiom),
  either
  \begin{itemize}
  \item From scratch.
    That is, commands that take as arguments
    a name and a type and/or a body,
    and that add
    the definition binding this name
    to this type and/or body.
    The canonical examples are
    \begin{lstlisting}[language=Coq]
Definition Name : type := body
    \end{lstlisting}

and

\begin{lstlisting}[language=Coq]
Axiom Name : type
\end{lstlisting}

    The type can also be given implicitly
    as the inferred type of the body,
    as in
    \begin{lstlisting}[language=Coq]
Definition Name := body
    \end{lstlisting}

  \item Saving the current (completely proven) theorem
    in the environment.
    These are
    the ``end of proof'' commands,
    such as \texttt{Qed}, \texttt{Save}, \texttt{Defined}.
  \end{itemize}
\item Commands that make progress
  in the current proof,
  which is necessarily made in several steps:
  \begin{enumerate}
  \item Opening a new theorem,
    as in
    \lstset{numbers=none}\begin{lstlisting}[language=Coq]
Theorem Name : type
    \end{lstlisting}

    or
    \begin{lstlisting}[language=Coq]
Definition Name : type
    \end{lstlisting}

  \item An arbitrary strictly positive
    amount of proof steps.
  \item Saving that theorem in the environment.
  \end{enumerate}
  These commands update
  (by adding exactly \emph{one} node)
  the internal \coq structure called
  ``proof tree''.
\item Commands that open a new theorem,
  that will be proven in multiple steps.
\end{enumerate}
The dependency tracking
is implemented as suitable hooks in the \coq functions
that the three kinds of commands eventually call.
When a new construct is registered in the environment,
the dependency tracking
walks over
the type
and body (if present)
of the new construct
and collects all constructs that are referenced.
When a  proof tree is updated,
the dependency tracking
examines
the top node
of the new proof tree
(note that this is always
the only change
with regards to the previous proof tree).
The commands
that update the proof tree
(that is, make a step in the current proof)
are called \texttt{tactics}.
\coq's tactic interpretation goes through three main phases:
\begin{enumerate}
\item parsing;
\item Ltac\footnote{Ltac is the \coq's tactical language,
used to combine tactics and add new user-defined tactics.} expansion;
\item evaluation.
\end{enumerate}
The tactic structure after each of these phases
is stored in the proof tree.
This allows to collect all construct references
mentioned at any of these tree levels.
For example, if tactic \texttt{Foo T} is defined as
\lstset{numbers=none}\begin{lstlisting}[language=Coq]
try apply BolzanoWeierstrass;
solve [ T | auto ]
\end{lstlisting}
and the user invokes the tactic as \texttt{Foo FeitThompson},
then the first level will contain (in parsed form)
\texttt{Foo FeitThompson},
the second level will contain (in parsed form)
\begin{lstlisting}[language=Coq]
try apply BolzanoWeierstrass;
solve [ FeitThompson | auto ].}
\end{lstlisting}
and the third level can contain any of:
\begin{itemize}
\item \texttt{refine (BolzanoWeierstrass \dots)},
\item \texttt{refine (FeitThompson \dots)},
\item something else, if the proof was found by \texttt{auto}.
\end{itemize}
The third level typically contains only
a few of the basic atomic fundamental rules (tactics) applications,
such as \texttt{refine}, \texttt{intro}, \texttt{rename} or \texttt{convert},
and combinations thereof.

\subsection{Dependency availability, format, and protocol}
\coq supports several interaction protocols: the \texttt{coqtop}, \texttt{emacs} and \texttt{coq-interface} protocols.   
Dependency tracking is available
in the program implementing
the \texttt{coq-interface} protocol which is designed for machine interaction.
The dependency information is printed
in a special message
for each \emph{potentially progress-making command}
that can give rise to a dependency.
A \emph{potentially progress-making command} is
one whose purpose is to change \coq's state.
For example, the command \texttt{Print Foo},
which displays the previously loaded mathematical construct \texttt{Foo},
is not a potentially progress-making command\footnote{Thus,
although this commands obviously needs item \texttt{Foo}
to be defined to succeed,
the dependency tracking
does not output that information.
That is not a problem in practice
because such commands are usually
issued by a user interface
to treat an interactive user request
(for example ``show me item \texttt{Foo}''),
but are not saved into the script
that is saved on disk.
Even if they were saved into the script,
adding or removing them to (from, respectively) the script
does not change the semantics of the script.}.
Any tactic invocation is a potentially progress-making command.
For example, the tactic \texttt{auto} silently succeeds (without any effect)
if it does not completely solve the goal it is assigned to solve.
In that case,
although that particular invocation did not make any actual progress
in the proof,
\texttt{auto} is still considered a potentially progress-making command,
and the dependency tracking outputs the message
\texttt{``dependencies: (empty list)''}.
Other kinds of progress-making commands include, for example
notation declarations and morphisms registrations.
Some commands, although they change \coq's state, 
might not give rise to a dependency.
For example, the \texttt{Set Firstorder Depth} command,
taking only an integer argument,
changes the maximum depth at which
the \texttt{firstorder}
tactic will search for a proof.
For such a command,
no dependency message is output.

One command may give rise to several dependency messages,
when they change \coq's state in several different ways.
For example,
the \texttt{intuition} tactic\footnote{
  The intuition tactic is a decision procedure for intuitionistic propositional calculus
  based on the contraction-free sequent calculi LJT* of Roy Dyckhof,
  extended to hand over subgoals which it cannot solve to another tactic.
}
can, mainly for efficiency reasons,
construct an ad hoc lemma,
register it into the global environment
and then use that lemma to prove the goal it has been assigned to solve,
instead of introducing the ad hoc lemma as a local hypothesis through a cut.
This is mainly an optimization:
The ad hoc lemma is defined as ``opaque'',
meaning that the typechecking (proofchecking) algorithm
is not allowed to unfold the body (proof) of the lemma
when the lemma is invoked,
and thus won't spend any time doing so.
By contrast,
a local hypothesis is always ``transparent'',
and the typechecking algorithm
is allowed to unfold its body.
For the  purpose of dependency tracking this means
that \texttt{intuition}
makes \emph{two} conceptually different steps:
\begin{enumerate}
\item register a new global lemma, under a fresh name;
\item solve the current subgoal in the proof currently in progress.
\end{enumerate}
Each of these steps gives rise to different dependencies.
For example, if the current proof is
\texttt{BolzanoWeierstrass},
then the new global lemma gives rise to dependencies of the form
\begin{quote}
  ``\texttt{BolzanoWeierstrass\_subproofN} depends on \dots''
\end{quote}
where the \texttt{\_subproofN} suffix is \coq's way of generating
a fresh name.
Closing of the subgoal by use of \texttt{BolzanoWeierstrass\_subproofN}
then gives rise to the dependency
\begin{quote}
  ``\texttt{BolzanoWeierstrass} depends on
  \texttt{BolzanoWeierstrass\_subproofN}''
\end{quote}

\subsection{Coverage and limitations}
\label{sec:limitations}

The \coq dependency tracking is already quite extensive, and
sufficient for
the whole Nijmegen \corn corpus.
Some restrictions remain
in parts of the \coq internals that the second author
does not yet fully understand.\footnote{
Such as when and how dynamics are used in tactic expressions
or a complete overview of all datatype tactics take as arguments.}
Our interests (and experiments) include not only purely mathematical dependencies
that can be found in the proof terms (for previous work see also \cite{Pons98,AspertiPCGS03}),
but also fast recompilation modes
for easy authoring of formal mathematics in large libraries and formal wikis.
The \coq dependency tracking code
currently finds all logically relevant dependencies from the proof terms,
even those that arise from automation tactics.
It does not handle yet the non-logical dependencies.
Examples include notation declarations,
morphism and equivalence relation declarations,\footnote{So
that the tactics for equality can handle one's user-defined equality.}
\texttt{auto} hint database registrations,\footnote{\texttt{auto} not only needs that the necessary lemmas be
available in the environment, but it also needs to be specifically instructed
to try to use them, through a mechanism where the lemmas are registered
in a ``hint database''. Each invocation of \texttt{auto} can specify
which hint databases to use.} but also tactic interpretation.
At this stage, we don't handle most of these,
but as already explained,
the internal structure of \coq lends itself well
to collecting dependencies that appear at the various levels
of tactic interpretation.
This means that we can already handle the (\emph{non-semantic}) dependencies
on logical constructs
that appear during the tactic interpretation,
but that do not end up being used in the final proof term.

Some of the non-logical dependencies are a more difficult issue.
For example, several dependencies related to tactic parametrization
(\texttt{auto} hint databases, \texttt{firstorder} proof depth search)
need specific knowledge of how the tactic is influenced by parameters,
or information available only to the internals of the tactic.
The best approach to handle such dependencies
seems to be to change (at the \ocaml source level in \coq)
the type of a tactic, so that the tactic itself is responsible
for providing such dependencies.
This will however have to be validated in practice, provided that we
manage to persuade the greater \coq community about the importance and
practical usefulness of complete dependency tracking for formal
mathematics and for research based on it.

\coq also presents an interesting corner case as far as opacity
of dependencies is concerned.
On the one hand, \coq has an explicit management of opacity of items;
an item originally declared as opaque
can only be used generically with regards to its type;
no information arising from its body can be used,
the only information available to other items
is the type.
Lemmas and theorems are usually declared opaque\footnote{thereby
following the mathematical principle of \texttt{proof irrelevance}.},
and definitions usually declared transparent,
but this is not forced by the system.
In some cases, applications of lemmas
need to be transparent.
\coq provides an easy way to decide
whether a dependency is opaque or transparent:
dependencies on opaque objects can only be opaque,
and dependencies on transparent objects are to be considered transparent.

Note that the predicative calculus of inductive constructions (pCIC)
uses a universe level structure,
where the universes have to be ordered in a well-founded way
at all times.
However, the ordering constraints between the universes are
hidden from the user,
and are absent from the types (statements) the user writes.
Changing the proof of a theorem \mT
can potentially have an influence on the universe constraints
of the theorem.
Thus, changing the body of an opaque item \mTp appearing in the proof
of \mT
can change the universe constraints
attached to it,
potentially in a way that is incompatible
with the way it was previously used
in the body of \mT.
Detecting whether the universe constraints
have changed or not
is not completely straightforward,
and needs specific knowledge
of the pCIC.
But unless one does so,
for complete certainty of correctness
of the library as a whole,
one has to consider \emph{all} dependencies as transparent.
Note that in practice
universe constraint incompatibilities are quite rare.
A large library may thus optimize its rechecking
after a small change, and not immediately follow opaque reverse dependencies.
Instead, fully correct universe constraint checking
could be done in a postponed way,
for example by rechecking the whole library from scratch
once per week or per month.

\section{Dependency extraction in \mizar}\label{sec:mizar}

Dependency computation in \mizar differs from the implementation
provided for \coq, being in some sense much simpler, but at the same
time also more robust with respect to the potential future
changes of the \mizar codebase. For comparison of the techniques, see
Section~\ref{sec:dep-summary}.  
For a more
detailed discussion of \mizar, see~\cite{mizar-first-30} or~\cite{mizar-in-a-nutshell}.

In \mizar{}, every article $A$ has its own environment
$\mathcal{E}_{A}$ specifying the context (theorems, definitions,
notations, etc.) that is used to verify the article.
$\mathcal{E}_{A}$, is usually a rather conservative overestimate of
the items that the article actually needs.  For example, even if an
article $A$ needs only one definition (or theorem, or notation, or
scheme, or\dots) from article $B$, all the definitions (theorems,
notations, schemes, \dots) from $B$ will be present in
$\mathcal{E}_{A}$.  The \emph{dependencies for an article $A$} are computed as the smallest environment
$\mathcal{E}_{A}^{\prime}$ under which $A$ is still \mizar-verifiable
(and has the same semantics as $A$ did under $\mathcal{E}_{A}$). To get 
dependencies of a particular \mizar item $I$ (theorem, definition, etc.,), we first
create a \emph{microarticle} containing essentially just the item $I$, and compute 
the dependencies of this microarticle.

More precisely, computing fine-grained dependencies in \mizar takes
three steps:
\begin{description}
\item[Normalization] Rewrite every article of the \mizar Mathematical
  Library so that:
  \begin{itemize}
  \item Each definition block defines exactly one concept.

    Definition blocks that contain multiple definitions or notations
    can lead to false positive dependencies.  For example, if two
    functions $g$ and $g$ are defined in a single definition block,
    and a theorem $\phi$ uses $f$ but not $g$, then we want to be able
    to say that $\phi$ depends on $f$ but is independent of $g$.
    Without splitting definition blocks, we have the specious
    dependency of $\phi$ upon $g$.
  \item All toplevel logical linking is replaced by explicit
    reference: constructions such as
    \begin{lstlisting}[language=Mizar]
@$\phi$@; then @$\psi$@;
    \end{lstlisting}
    whereby the statement $\psi$ is justified by the statement $\phi$,
    are replaced by
  \begin{lstlisting}[language=Mizar]
Label1: @$\phi$@;
Label2: @$\psi$@ by Label1;
  \end{lstlisting}
  where \verb+Label1+ and \verb+Label2+ are new labels.  By doing this
  transformation, we ensure that the only way that a statement is
  justified by another is through explicit reference.
\item Segments of reserved variables all have length exactly $1$.  For
  example, constructions such as
    \begin{lstlisting}[language=Mizar]
reserve A for set,
        B for non empty set,
        f for Function of A, B,
        M for Cardinal;
    \end{lstlisting}
    which is a single reservation statement that assigns types to four
    variables (\verb+A+, \verb+B+, \verb+f+, and \verb+M+) is replaced
    by four reservation statements, each of which assigns a type to a
    single variable:
    \begin{lstlisting}[language=Mizar]
reserve A for set;
reserve B for non empty set;
reserve f for Function of A, B;
reserve M for Cardinal;
    \end{lstlisting}
    When reserved variables are normalized in this way, one can
    eliminate some false positive dependencies.  A theorem in which,
    say, the variable \verb+f+ occurs freely but which has nothing to
    do with cardinal numbers has the type \verb+Function of A,B+ in
    the presence of both the first and the second sequences of
    reserved variables.  If the first reservation statement is
    deleted, the theorem becomes ill-formed because \verb+f+ no longer
    has a type.  But the reservation statement itself directly
    requires that the type \verb+Cardinal+ of cardinal numbers is
    available, and thus indirectly requires a part of the development
    of cardinal numbers.  If the theorem has nothing to do with
    cardinal numbers, this dependency is clearly specious.  By rewriting
    reserved variables in the second way, though, one sees that one
    can safely delete the fourth reservation statement, thereby
    eliminating this false dependency.
  \end{itemize}
  These rewritings do not affect the semantics of the \mizar article.
\item[Decomposition] For every normalized article $A$ in the \mizar
  Mathematical Library, extract the sequence $\langle I_{1}^{A},
  I_{2}^{A}, \dots , I_{n}^{A} \rangle$ of its toplevel items, each of
  which written to a ``microarticle'' $A_{k}$ that contains only
  $I_{k}^{A}$ and whose environment is that of $A$ and contains each
  $A_{j}$ ($j < k$).
\item[Minimization] For every article $A$ of the \mizar Mathematical
  Library and every microarticle $A_{n}$ of $A$, do a brute-force
  minimization of smallest environment $\mathcal{E}_{A_{n}}$ such that
  $A_{n}$ is \mizar{}-verifiable.
\end{description}

The brute-force minimization works as follows.  Given a microarticle
$A$, we successively trim the environment for all the \mizar{}
environment item kinds.\footnote{Namely, theorems, schemes, top-level
  lemmas, definitional theorems, definientia, patterns, registrations,
  and constructors.  See~\cite{mizar-in-a-nutshell} for a discussion
  of these item kinds.} Each item kind is associated with a sequence
$s$ of imported items $\langle a_{1}, \dots,$ $a_{n} \rangle$ and the
task is to find a minimal sublist $s^{\prime}$ of $s$ such that $A$ is
\mizar{}-verifiable.\footnote{There is always one minimal sublist,
  since we assume that $A$ is \mizar{}-verifiable to begin with.
} We apply a simple binary search algorithm to $s$ to compute the
minimal sublist $s^{\prime}$.  Applying this approach for all \mizar{}
item kinds, for all microarticles $A_{k}$, for all articles $A$ of the
\mizar Mathematical Library is a rather expensive computation (for
some \mizar{} articles, this process can take several hours).  It is
much slower than the method used for \coq described in
Section~\ref{sec:coq}.  However the result is truly minimized, which
is important for many applications of dependencies.  Additionally, we
have already developed some heuristics that help to find $s^{\prime}$,
and these already do perform tolerably fast.

\section{Comparison of the Methods}\label{sec:dep-summary}

Some observations comparing the \coq and \mizar dependency
computation can be drawn generally, without comparing the actual data
as done in the following sections.
Dependencies in the case of \corn{} are generated by hooking
into the actual code and are thus quite exactly mirroring the work of the proof assistant.

In the case of \mizar{}, dependencies are approximated from above.  The
dependency graph in this case starts with an over-approximation of what
is known to be sufficient for an item to be \mizar{}-verifiable and
then successively refines this over-approximation toward a minimal set
of sufficient conditions.  
A significant difference is that the dependencies in \coq{} are not
minimized: the dependency tracking there tells us exactly the
dependencies that were used by \coq{} (in the particular context) when
a certain command is run. Thus, if for example the context is rich,
and redundant dependencies are used by some exhaustive strategies, we
will not detect their redundancy. On the other hand, in \mizar{} we do
not rely on the proof assistant reporting how it exactly works, and
instead try to exhaustively minimize the set of dependencies, until an
error occurs. This process is more computationally intensive, however,
it guarantees minimality (relative to the proof assistant's power)
which is interesting for many of the applications mentioned below.

Another difference is in the coverage of non-logical
constructs. Practically every resource necessary for a verification of
a \mizar article is an explicit part of the article's
environment. Thus, it is easy to minimize not just the strictly
logical dependencies, but also the non-logical ones, like the sets of
symbols and notations needed for a particular item, or particular
automations like definitional expansions. For LCF-based proof
assistants, this typically implies further work on the dependency
tracking.

\section{Evaluation, Experiments, and Applications}\label{Experiments-and-Results}

\subsection{Dependency extraction for \corn and \mml}
\label{dep-stats}
We use the dependency extraction methods described
in~\ref{sec:coq} and~\ref{sec:mizar} to obtain fine dependency data for
the \corn library and an initial 100 article fragment of the \mml.
As described above, for
\corn, we use the dependency exporter implemented directly using the
\coq code base. The export is thus approximately as fast as the \coq
processing of \corn itself, taking about 40 minutes on contemporary
hardware. The product are for each \corn file a corresponding file
with dependencies, which have altogether about 65 MB. This
information is then post-processed by standard UNIX and other tools into the
dependency graph discussed below.

For \mizar and \mml we use the brute-force dependency extraction
approach discussed above. This takes significantly longer than \mizar
processing alone, also because of the number of preprocessing and
normalization steps that need to be done when splitting articles into
micro-articles. For our data this now takes about one day for the
initial 100 article fragment of the \mml, the main share of this time
being spent on minimizing the large numbers of items used implicitly
by \mizar. Note that in this implementation we are initially more
interested in achieving completeness and minimality rather than efficiency, and a
number of available optimizations can reduce this time
significantly\footnote{For example a very simple recent optimization
  done for theorems, definitions, and schemes, has reduced the
  processing time in half.}. The data obtained are again post-processed
by standard UNIX tools into the dependency graphs.

In order to compare the benefits of having fine dependencies, we also
compute for each library the \textit{full file-based dependency} graph
for all items. These graphs emulate the current dumb file-based
treatment of dependencies in these libraries: each time an item is
changed in some file, all items in the depending files have to be
re-verified. The two kinds of graphs for both libraries are then compared in Table~\ref{tab:stat}.

The graphs confirm our initial intuition that having the fine
dependencies will significantly speed up partial recompilation of the
large libraries, which is especially interesting in the \corn and \mml
formal wikis that we
develop.\footnote{\url{http://mws.cs.ru.nl/mwiki/},
  \url{http://mws.cs.ru.nl/cwiki/}} For example, the average number of
items that need to be recompiled when a random item is changed has
dropped about seven times for \corn, and about five times for
\mizar. The medians for these numbers are even more interesting,
increasing to fifteen for \mizar. The difference between \mml and
\corn is also quite interesting, but it is hard 
to draw any conclusions. The corpora differ in their 
content and use different styles and techniques.

\begin{table}[htb]
  \centering
  \begin{tabular}{@{\extracolsep{0.3cm}}l|*{4}{D{.}{.}{8.1}}}
    &\multicolumn{1}{c}{CoRN/item}&\multicolumn{1}{c}{CoRN/file}&\multicolumn{1}{c}{MML-100/item}&\multicolumn{1}{c}{MML-100/file}\\
    \hline
    Items& 9\:462 & 9\:462 & 9\:553 & 9\:553\\
    Deps& 175\:407 &2\:214\:396& 704\:513 & 21\:082\:287\\
    TDeps\phantom{j}& 3\:614\:445&24\:385\:358& 7\:258\:546 & 34\:974\:804\\
    P(\%)&8 & 54.5& 15.9 & 76.7 \\
    ARL&382 & 2\:577.2 & 759.8 & 3\:661.1\\
    MRL& 12.5 & 1\:183 & 155.5 & 2\:377.5\\
  \end{tabular}\\
  {\small
  \begin{description}
  \item[Deps] Number of dependency edges
  \item[TDeps] Number of transitive dependency edges
  \item[P] Probability that given two randomly chosen items,
    one depends (directly or indirectly) on the other, or vice versa.
  \item[ARL] Average number of items recompiled if one item is changed.
  \item[MRL] Median number of items recompiled if one item is changed.
  \end{description}}
  \caption{Statistics of the item-based and file-based dependencies for \corn and \mml}
  \label{tab:stat}
\end{table}
\vspace{-17mm}
\begin{figure}[htbp]
  \centering
  \input{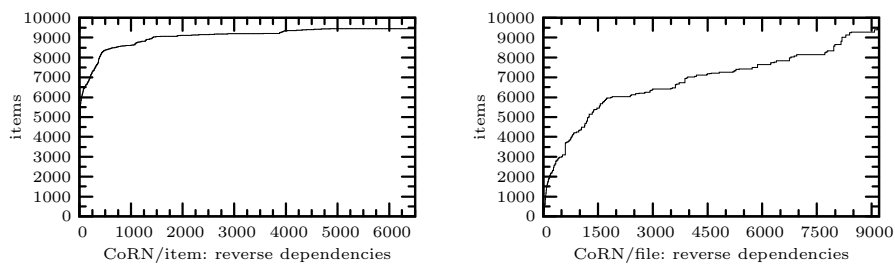}
\setlength{\unitlength}{0.120450pt}
\begin{picture}(1380,828)(0,0)
\fontsize{6}{7.68246}\selectfont
\thicklines \path(253,156)(294,156)
\thicklines \path(1307,156)(1266,156)
\put(229,156){\makebox(0,0)[r]{ 0}}
\thicklines \path(253,187)(273,187)
\thicklines \path(1307,187)(1287,187)
\thicklines \path(253,218)(294,218)
\thicklines \path(1307,218)(1266,218)
\put(229,218){\makebox(0,0)[r]{ 1000}}
\thicklines \path(253,250)(273,250)
\thicklines \path(1307,250)(1287,250)
\thicklines \path(253,281)(294,281)
\thicklines \path(1307,281)(1266,281)
\put(229,281){\makebox(0,0)[r]{ 2000}}
\thicklines \path(253,312)(273,312)
\thicklines \path(1307,312)(1287,312)
\thicklines \path(253,343)(294,343)
\thicklines \path(1307,343)(1266,343)
\put(229,343){\makebox(0,0)[r]{ 3000}}
\thicklines \path(253,374)(273,374)
\thicklines \path(1307,374)(1287,374)
\thicklines \path(253,406)(294,406)
\thicklines \path(1307,406)(1266,406)
\put(229,406){\makebox(0,0)[r]{ 4000}}
\thicklines \path(253,437)(273,437)
\thicklines \path(1307,437)(1287,437)
\thicklines \path(253,468)(294,468)
\thicklines \path(1307,468)(1266,468)
\put(229,468){\makebox(0,0)[r]{ 5000}}
\thicklines \path(253,499)(273,499)
\thicklines \path(1307,499)(1287,499)
\thicklines \path(253,530)(294,530)
\thicklines \path(1307,530)(1266,530)
\put(229,530){\makebox(0,0)[r]{ 6000}}
\thicklines \path(253,562)(273,562)
\thicklines \path(1307,562)(1287,562)
\thicklines \path(253,593)(294,593)
\thicklines \path(1307,593)(1266,593)
\put(229,593){\makebox(0,0)[r]{ 7000}}
\thicklines \path(253,624)(273,624)
\thicklines \path(1307,624)(1287,624)
\thicklines \path(253,655)(294,655)
\thicklines \path(1307,655)(1266,655)
\put(229,655){\makebox(0,0)[r]{ 8000}}
\thicklines \path(253,686)(273,686)
\thicklines \path(1307,686)(1287,686)
\thicklines \path(253,718)(294,718)
\thicklines \path(1307,718)(1266,718)
\put(229,718){\makebox(0,0)[r]{ 9000}}
\thicklines \path(253,749)(273,749)
\thicklines \path(1307,749)(1287,749)
\thicklines \path(253,780)(294,780)
\thicklines \path(1307,780)(1266,780)
\put(229,780){\makebox(0,0)[r]{ 10000}}
\thicklines \path(253,156)(253,197)
\thicklines \path(253,780)(253,739)
\put(253,107){\makebox(0,0){ 0}}
\thicklines \path(310,156)(310,176)
\thicklines \path(310,780)(310,760)
\thicklines \path(368,156)(368,176)
\thicklines \path(368,780)(368,760)
\thicklines \path(425,156)(425,197)
\thicklines \path(425,780)(425,739)
\put(425,107){\makebox(0,0){ 1500}}
\thicklines \path(482,156)(482,176)
\thicklines \path(482,780)(482,760)
\thicklines \path(539,156)(539,176)
\thicklines \path(539,780)(539,760)
\thicklines \path(597,156)(597,197)
\thicklines \path(597,780)(597,739)
\put(597,107){\makebox(0,0){ 3000}}
\thicklines \path(654,156)(654,176)
\thicklines \path(654,780)(654,760)
\thicklines \path(711,156)(711,176)
\thicklines \path(711,780)(711,760)
\thicklines \path(769,156)(769,197)
\thicklines \path(769,780)(769,739)
\put(769,107){\makebox(0,0){ 4500}}
\thicklines \path(826,156)(826,176)
\thicklines \path(826,780)(826,760)
\thicklines \path(883,156)(883,176)
\thicklines \path(883,780)(883,760)
\thicklines \path(940,156)(940,197)
\thicklines \path(940,780)(940,739)
\put(940,107){\makebox(0,0){ 6000}}
\thicklines \path(998,156)(998,176)
\thicklines \path(998,780)(998,760)
\thicklines \path(1055,156)(1055,176)
\thicklines \path(1055,780)(1055,760)
\thicklines \path(1112,156)(1112,197)
\thicklines \path(1112,780)(1112,739)
\put(1112,107){\makebox(0,0){ 7500}}
\thicklines \path(1170,156)(1170,176)
\thicklines \path(1170,780)(1170,760)
\thicklines \path(1227,156)(1227,176)
\thicklines \path(1227,780)(1227,760)
\thicklines \path(1284,156)(1284,197)
\thicklines \path(1284,780)(1284,739)
\put(1284,107){\makebox(0,0){ 9000}}
\thicklines \path(253,780)(253,156)(1307,156)(1307,780)(253,780)
\put(36,468){\makebox(0,0)[l]{\rotatebox[origin=c]{90}{items}}}
\put(780,34){\makebox(0,0){CoRN/file: reverse dependencies}}
\thinlines \path(253,156)(253,156)(253,156)(253,156)(253,156)(253,156)(253,157)(253,157)(253,157)(254,157)(254,157)(254,157)(254,158)(254,158)(254,159)(254,159)(254,160)(254,160)(254,160)(254,160)(254,161)(254,161)(254,161)(254,161)(254,162)(255,162)(255,163)(255,163)(255,166)(255,166)(255,166)(255,166)(255,167)(255,167)(255,169)(255,169)(255,172)(256,172)(256,173)(256,173)(256,173)(256,173)(256,175)(256,175)(256,177)(256,177)(256,178)(256,178)(256,180)(257,180)(257,180)
\thinlines \path(257,180)(257,180)(257,180)(257,180)(257,181)(257,181)(257,183)(257,183)(257,186)(258,186)(258,187)(258,187)(258,190)(258,190)(258,191)(258,191)(258,192)(258,192)(258,195)(259,195)(259,201)(259,201)(259,205)(259,205)(259,208)(259,208)(259,210)(260,210)(260,211)(260,211)(260,215)(260,215)(260,218)(260,218)(260,219)(261,219)(261,222)(261,222)(261,222)(261,222)(261,223)(261,223)(261,229)(261,229)(261,229)(262,229)(262,235)(262,235)(262,238)(262,238)(262,238)
\thinlines \path(262,238)(263,238)(263,242)(263,242)(263,246)(263,246)(263,247)(264,247)(264,249)(264,249)(264,251)(265,251)(265,254)(265,254)(265,254)(265,254)(265,256)(266,256)(266,258)(266,258)(266,258)(266,258)(266,260)(267,260)(267,260)(267,260)(267,264)(268,264)(268,266)(269,266)(269,270)(269,270)(269,271)(270,271)(270,274)(271,274)(271,274)(271,274)(271,275)(272,275)(272,277)(273,277)(273,283)(274,283)(274,286)(276,286)(276,291)(277,291)(277,292)(279,292)(279,293)
\thinlines \path(279,293)(281,293)(281,297)(282,297)(282,300)(284,300)(284,301)(285,301)(285,303)(285,303)(285,308)(285,308)(285,310)(287,310)(287,311)(287,311)(287,315)(287,315)(287,316)(288,316)(288,317)(290,317)(290,318)(290,318)(290,321)(292,321)(292,324)(292,324)(292,329)(293,329)(293,330)(296,330)(296,333)(299,333)(299,337)(301,337)(301,339)(302,339)(302,340)(304,340)(304,342)(311,342)(311,342)(311,342)(311,344)(313,344)(313,349)(322,349)(322,386)(323,386)(323,387)
\thinlines \path(323,387)(325,387)(325,388)(325,388)(325,388)(326,388)(326,388)(326,388)(326,389)(327,389)(327,389)(327,389)(327,389)(327,389)(327,389)(328,389)(328,389)(328,389)(328,390)(328,390)(328,390)(332,390)(332,392)(334,392)(334,395)(335,395)(335,396)(339,396)(339,399)(339,399)(339,402)(341,402)(341,405)(343,405)(343,407)(343,407)(343,407)(343,407)(343,409)(345,409)(345,413)(348,413)(348,416)(349,416)(349,419)(350,419)(350,419)(357,419)(357,421)(364,421)(364,422)
\thinlines \path(364,422)(364,422)(364,426)(367,426)(367,427)(372,427)(372,431)(373,431)(373,436)(383,436)(383,442)(384,442)(384,448)(388,448)(388,451)(389,451)(389,455)(389,455)(389,455)(391,455)(391,458)(392,458)(392,462)(392,462)(392,466)(396,466)(396,469)(397,469)(397,473)(397,473)(397,477)(407,477)(407,483)(409,483)(409,485)(411,485)(411,487)(411,487)(411,490)(411,490)(411,491)(415,491)(415,492)(421,492)(421,496)(429,496)(429,503)(432,503)(432,505)(432,505)(432,505)
\thinlines \path(432,505)(432,505)(432,510)(439,510)(439,515)(443,515)(443,517)(445,517)(445,520)(448,520)(448,522)(448,522)(448,523)(450,523)(450,524)(451,524)(451,526)(452,526)(452,527)(453,527)(453,528)(465,528)(465,530)(469,530)(469,532)(526,532)(526,536)(529,536)(529,538)(543,538)(543,542)(553,542)(553,543)(571,543)(571,545)(573,545)(573,546)(588,546)(588,551)(590,551)(590,552)(595,552)(595,554)(595,554)(595,555)(597,555)(597,556)(654,556)(654,558)(658,558)(658,560)
\thinlines \path(658,560)(668,560)(668,567)(668,567)(668,570)(672,570)(672,572)(680,572)(680,576)(698,576)(698,589)(702,589)(702,591)(706,591)(706,594)(724,594)(724,594)(728,594)(728,596)(734,596)(734,600)(747,600)(747,600)(766,600)(766,601)(768,601)(768,604)(776,604)(776,605)(777,605)(777,605)(781,605)(781,606)(806,606)(806,609)(847,609)(847,609)(847,609)(847,609)(848,609)(848,609)(848,609)(848,610)(848,610)(848,610)(849,610)(849,610)(850,610)(850,610)(851,610)(851,611)
\thinlines \path(851,611)(851,611)(851,611)(851,611)(851,611)(852,611)(852,612)(852,612)(852,613)(853,613)(853,613)(855,613)(855,614)(858,614)(858,618)(868,618)(868,619)(909,619)(909,624)(926,624)(926,633)(964,633)(964,633)(967,633)(967,635)(967,635)(967,637)(968,637)(968,639)(978,639)(978,641)(985,641)(985,645)(1026,645)(1026,654)(1031,654)(1031,657)(1041,657)(1041,660)(1049,660)(1049,664)(1137,664)(1137,670)(1138,670)(1138,670)(1140,670)(1140,672)(1144,672)(1144,673)(1144,673)(1144,675)
\thinlines \path(1144,675)(1146,675)(1146,676)(1165,676)(1165,691)(1169,691)(1169,696)(1188,696)(1188,706)(1191,706)(1191,719)(1203,719)(1203,726)(1215,726)(1215,733)(1220,733)(1220,735)(1293,735)(1293,742)(1302,742)(1302,746)(1307,746)
\thicklines \path(253,780)(253,156)(1307,156)(1307,780)(253,780)
\end{picture}
  \caption{Cumulative transitive reverse dependencies for \corn: file-based vs.\ item-based}
  \label{fig:CoRN_cumul}
\end{figure}
\begin{figure}[htbp]
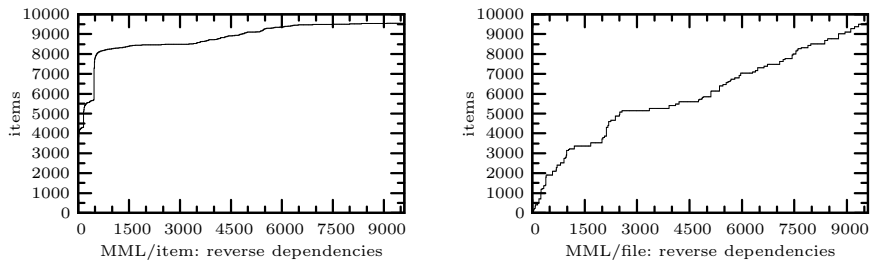

  \centering
  \input{MML_finedeps.eepic}
  \input{MML_filedeps.eepic}
  \caption{Cumulative transitive reverse dependencies for \mml: file-based vs.\ item-based}
  \label{fig:MML_cumul}
\end{figure}
Another interesting new statistics given in Table~\ref{Links} is the
information about the number and structure of \textit{explicit} and
\textit{implicit} dependencies that we have done for \mizar. Explicit
dependencies are anything that is already mentioned in the original
text.
Implicit dependencies are everything else, for example
dependencies on type mechanisms. 
Note that the ratio
of implicit dependencies is very significant, which suggests that
handling them precisely can be quite necessary for the learning and
ATP experiments conducted in the next section.

\begin{table*}[htbp]
\begin{center}
  \begin{tabular}{|l|r|r|r|r|r|}
    \hline
             &theorem & top-level lemma & definition & scheme & registration \\
    \hline
from & 550134 & 44120 & 44216 & 7053 & 58622 \\
    \hline
to & 314487 & 2384 & 263486 & 6510 & 108449 \\
    \hline
  \end{tabular}
\end{center}
\label{Links}
  \caption{Statistics of Mizar direct dependencies from and to different items}
\end{table*}

\subsection{Dependency analysis for AI-based proof assistance}
\label{sec:mach-learn-appl}
The knowledge of how a large number of theorems are proved is used by
mathematicians to direct their new proof attempts and theory
developments.  In the same way, the precise formal proof knowledge that we
now have can be used for directing formal automated theorem proving (ATP)
systems and meta-systems over the large mathematical
libraries. In~\cite{AlamaKTUH11} we provide an initial evaluation of
the usefulness of our \mml dependency data for machine learning of such proof guidance of first-order ATPs.

These experiments are conducted on a set of 2078 problems extracted
from the \mizar library and translated 
to
first-order ATP format. We emulate the growth of the \mizar library (limited
to the 2078 problems), by considering all previous theorems and
definitions when a new conjecture is attempted (i.e., when a new theorem is
formulated by an author, requiring a proof).  The ATP problems thus
become very large, containing thousands of the previously proved
formulas as available axioms, which obviously makes automated theorem
proving quite difficult, see e.g.~\cite{UrbanHV10} and~\cite{MengP09}
for details. We run the state-of-the-art \Vampire{}-SInE~\cite{HoderV11} ATP system on
these large problems, and solve 567 of them (with a 10-second
timelimit). Then, instead of attacking such large problems directly, we learn
proof relevance from all previous fine-grained proof dependencies, using machine
learning with a naive Bayes classifier.
This technique works surprisingly well: in comparison with running
\Vampire{}-SInE directly on the large problems, the problems pruned by
such trained machine learner can be proved by \Vampire{} in 717 cases,
i.e., the efficiency of the automated theorem proving is raised by
about 30\% when we apply the knowledge about previous proof
dependencies, which is a very significant advance in the
world of automated theorem proving, where the search complexity is
typically superexponential. 

In~\cite{AlamaKU12} we further leverage 
this automated reasoning technique by scaling the dependency analysis 
to the whole \MML, and attempting a fully automated proof for every \MML theorem.
This yields the so-far largest number of fully automated proofs over the whole \MML, 
allowing us (using the precise formal dependencies of the ATP and \MML proofs)
 to attempt an initial comparison of human and automated proofs in general mathematics.

\subsection{Interactive editing with fine-grained dependencies}
\label{sec:interactive-editor}
A particular practical use of fine dependencies (initially motivating
the work done on \coq dependencies in~\ref{sec:coq}) is for advanced interactive editing.
\tmegg \cite{LEM_tmegg} is a \TeXmacs-based
user interface to \coq.\footnote{
The dependency tracking for \coq was actually started by the second author as
 part of the development of \tmegg.
This 
facility has been already integrated
in the official release of \coq.
Since then this facility was extended to be able to treat the whole of
the \corn library. These changes are not yet included in the official release of \coq.}
Its main purpose is to integrate formal mathematics
done in \coq in a more general document
(such as course notes or journal article)
without forcing the document to follow the structure of
the formal mathematics contained therein.

For example,
it does not require
that the order in which the mathematical constructs
appear in the document
be the same as the order in which they are presented to \coq.
As one would expect, the latter must respect the constraints
inherent to the incremental construction of the formal mathematics,
such as
a lemma being proven before it is used in the proof of a theorem
or
a definition being made before the defined construct is used.

However,
the presentation the author would like to put in the document
may not strictly respect these constraints.
For example,
clarity of exposition may benefit from
first presenting the proof of the main theorem,
making it clear how each lemma being used is useful,
and then only go through all lemmas.
Or a didactic presentation of a subject may
first want to go through some examples
before presenting the full definitions
for the concepts being manipulated.

\tmegg thus allows the mathematical constructs to be
in any order in the document, and
uses the dependency information to dynamically
---~and lazily~---
load any construct necessary to perform the requested action.
For example, if the requested action is
``check the proof of this theorem'',
it will automatically load all definitions and lemmas
used by the statement or proof of the theorem.

An interactive editor presents slightly different requirements
than the batch recompilation scenario of a mathematical library
described in~\ref{dep-stats}.
One such difference is that
an interactive editor needs the dependency information,
as part of the interactive session,
for partial in-progress proofs.
Indeed,
if any in-progress proof depends on an item \mT,
and the user wishes to change or unload (remove from the environment) \mT,
then the part of the in-progress proof
that depends on \mT has to be undone,
even if the dependency is opaque.

\section{Related Work}
\label{Related}
Related work exists in the first-order ATP field, where a number of
systems can today output the axioms needed for a particular proof. 
Purely semantic (proof object) dependencies have been extracted several times for
several ITPs, for example by Bertot and the Helm project for
\coq~\cite{Bertot00dependencygraphs,Pons98,AspertiPCGS03}, and Obua and McLaughlin
for HOL Light and \isabelle. The focus of the latter two dependency
extractions is on cross-verification, and are based on quite low-level
(proof object) mechanisms.
A higher-level\footnote{By \emph{higher-level} we mean tracking
  \emph{higher-level} constructs, like use of theorems and tactics,
  not just tracking of the low-level primitive steps done in the
  proof-assistant's kernel.} semantic dependency exporter for HOL
Light was recently implemented by Adams~\cite{Adams-icms} for his work
on HOL Light re-verification in HOL Zero. This could be usable as a
basis for 
extending our applications to the core HOL Light library and the
related large Flyspeck library. The \coq/\corn{} approach quite likely
easily scales to other large \coq{} libraries, like for example the
one developed in the Math Components project~\cite{Gonthier}. 
Our focus in this work is wider than the semantic-only
efforts: We attempt to get the full information about all implicit
mechanisms (including syntactic mechanisms), and we are interested in
using the information for smart re-compilation, which requires to
track much more than just the purely semantic or low-level
information.

\section{Conclusion and Future Work}\label{Vision}
In this paper we have tried to show the importance and attractiveness of
formal dependencies. We have implemented and used two very different techniques to elicit
fine-grained proof dependencies for two very different proof
assistants and two very different large formal mathematical libraries.
This provides enough confidence that our approaches will scale to
other important libraries and assistants, and our techniques and the
derived benefits will be usable in other contexts.

Mathematics is being increasingly encoded in a computer-understandable
(formal) and in-principle-verifiable way.  The results are
increasingly large interdependent computer-understandable libraries of
mathematical knowledge.  (Collaborative) development and refactoring
of such large libraries requires advanced computer support, providing
fast computation and analysis of dependencies, and fast
re-verification methods based on the dependency information.  
As such automated
assistance tools reach greater and greater reasoning power, the
cost/benefit ratio of doing formal mathematics decreases.

Given our previous work on several parts of this program, providing
exact dependency analysis and linking it to the other important tools
seems to be a straightforward choice. Even though the links to proof
automation, fast large-scale refactoring, and proof analysis, are very
fresh, it is our hope that the significant performance boosts already
sufficiently demonstrate the importance of good formal dependency
analysis for formal mathematics, and for future mathematics in
general.

\bibliographystyle{splncs03}
\bibliography{dependencies}

\end{document}